\documentclass{PoS}

\title{No Tau? No Astronomy!}

\ShortTitle{No \protect{$\tau$}? No \protect{$\nu$} Astronomy}

\author{\speaker{Daniele Fargion}\\
       Physics Department \& INFN Rome1, Rome University 1, P.le A. Moro 2, 00185, Rome, Italy\\
       MIFP, Via Appia Nuova 31, 00040 Marino (Rome), Italy\\
       E-mail: \email{daniele.fargion@roma1.infn.it}}

\author{Pietro Oliva\\
      Niccol\`o Cusano University, Via Don Carlo Gnocchi 3, 00166 Rome, Italy\\
      MIFP, Via Appia Nuova 31, 00040 Marino (Rome), Italy\\
      Department of Sciences, University Roma Tre, Via Vasca Navale 84, 00146 Rome, Italy\\
        E-mail: \email{pietro.oliva@unicusano.it}}

\FullConference{35th International Cosmic Ray Conference\\
	 10-20 July, 2017\\
	 Bexco, Busan, Korea}

\usepackage{journals, amsmath}

\abstract{
Since 2013 IceCube cascade showers sudden overabundance have shown a fast flavor change above 30-60 TeV up to PeV energy. This flavor change from dominant muon tracks at TeVs to shower events at higher energies, has been indebted to a new injection of a neutrino astronomy. However the recent published 54 neutrino HESE, high energy starting events, as well as the 38 external muon tracks made by trough going muon formed around the IceCube, none of them are pointing to any expected X-gamma or radio sources: no one in connection to GRB, no toward active BL Lac, neither to AGN source in Fermi catalog. No clear correlation with nearby mass distribution (Local Group), nor to galactic plane. Moreover there have not been any record (among a dozen of 200 TeV energetic events) of the expected double bang due to the tau neutrino birth and decay. An amazing and surprising unfair distribution in flavor  (suppressing tau) versus an expected democratic one. Finally there is not a complete consistence of the internal HESE event spectra and the external crossing muon track ones. Moreover the apparent sudden astrophysical neutrino flux rise at 60 TeV might be probably also suddenly cut at a few PeV in order to hide the (unobserved , yet) Glashow resonance peak at 6.3 PeV. A more mundane prompt charmed atmospheric neutrino component may explain most of the IceCube puzzles. If in this near future, 2017-2018, ICECUBE does not discover any tau neutrino signals somewhere (by double bang) there are a list of consequences to face. These missing correlations and in particular the tau signature absence force us to claim. as in a famous Martini spot: No Tau? No  Astronomy.
}

\begin{document}

\section{Introduction}
Since 2013 \cite{Fargion_2014} the sudden change of Ultra High Energy (UHE) neutrino flavor (IceCube high-energy starting events, HESE)  was discussed by us in the framework of the long-awaited birth of a UHE$\nu$ astronomy \cite{2017arXiv170200021F, 2017arXiv170500383K, 2015EPJWC..9908002F, 2014NuPhS.256..213F}.
Indeed, up to TeV energy the UHE$\nu$ are following the expected spectra of atmospheric $\nu$ secondaries for which it is found a flux ratio of $\nu_\mu$ ($\bar{\nu}_\mu$) over $\nu_e$ ($\bar{\nu}_e$) of $\phi_{\nu_\mu}:\phi_{\nu_e}\approx20:1$.
Suddenly, at few tens TeVs, the flavor revolution kicks in: four times more  shower or cascade events, started by $\nu_e$, or by $\nu_\tau$ or by Neutral Current (NC) events, (respect to muon tracks) appeared in IceCube:
$\phi_{\nu_\mu}^{\mathrm{tracks}}:(\phi_{\nu_e}+\phi_{\nu_\tau}+\phi_{\mathrm{NC}})=1:4$ \cite{2013arXiv1309.7003I}.

\begin{figure}[!b]
\begin{center}
\includegraphics[width=0.7\textwidth]{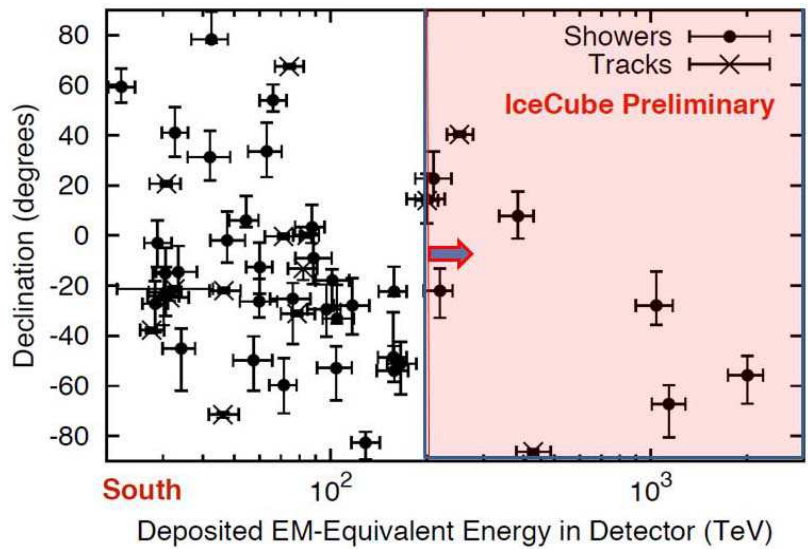}
\caption{About ten neutrino events above 200 TeV energy have been considered in the IceCube for the tau discover. Their absence is puzzling.
In near future, by double record times, by enlarged ICECUBE area and volume, by
enhanced filtering at lower energy threshold, the tau twin bangs, the events will double or multiply as they might be detected even at 100 TeVs energy.
} \label{Fig1}
\end{center}
\end{figure}

\begin{figure}[!t]
\begin{center}
\includegraphics[width=0.67\textwidth]{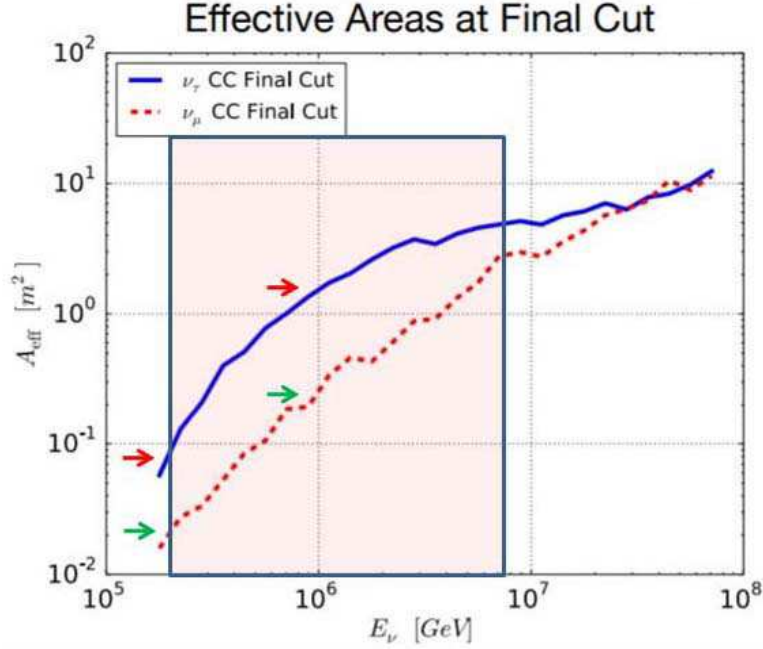}
\caption{Different effective area for the charged current among muon and tau neutrino. Note the remarkable difference and remind the presence of three HESE  muon tracks above 200 TeV while the more probable tau signal is still totally absent.
} \label{Fig2}
\end{center}
\end{figure}

Since 2015, IceCube observed an integral number of 54 HESE events where $\tau$ signals. among highest ten above $200$ TeV energy, are totally absent.
Unfortunately, no neutrino astronomy arose since then, though only very recently a high-energy $\nu$s flux at TeV energy made by cascades, claimed to be
weakly anisotropic were published \cite{2017arXiv170502383I}. We shall neglect it in present paper.
Indeed, above 200 TeV neutrino  energy the so-called double chain of events producing first a shower, $\nu_\tau$+N, and then  a consequent $\tau$ with its decay forward (ten meter or more) in a second shower (double bang mechanism) \cite{1995APh.....3..267L}, it was in principle observable but experimentally still undetected nor in 2013 nor in 2015 \cite{2016PhRvD..93b2001A}, neither it was in more recent analysis \cite{2017arXiv170205238X}.
In fact, at least a 10 of such events above 200~TeV were not revealing any $\tau$. Now, the $\nu_\mu$ and $\nu_e$, as known, are extremely polluted by atmospheric noises, while $\nu_\tau$ (and $\bar{\nu}_\tau$) are the unique beyond-doubt probe of an extraterrestrial signal. Why tau has not been detected? \cite{Fargion_WHY}
There are, truthfully, several arguments to keep in mind before any claim of $\nu$ astronomy arising from the 54 HESE events:\newline
1) There are no clear Galactic plane  signatures. Most gamma Fermi MeV-GeV and even TeV
 telescope astronomy do show  both rare extragalactic but also a clear galactic plane
signal. At energy below TeVs (tens GeV or below) the whole Universe is still transparent to  energy gamma sky: there is not any cosmic dominant signal over the galactic one. One of course may mention a PeV IceCube shower event pointing somehow to our Galactic Center direction but there are not too many additional clear galactic signals within the last 54 events. Let us mention also as an additional analogy, the Gamma Ray Bursts:
(GRB) are cosmic and isotropic but their twin Soft Gamma Repeaters (SGR) are mostly galactic and they  present in a non negligible number.\\
2) There are no GRB-$\nu$ correlated events. These GRBS events are the most (apparent) gamma brightest flashes and they were expected to be in correlated shining activity with highest neutrino energy (if these GRBS sources are of hadronic nature). A possible solution we suggested occurs in a different scenario for GRB sources made mostly by binary BH-NS collapse forming a jet not by hadrons jets but just by electron pairs jets, see \cite{2016arXiv1605.00177}.\\
3) There are no at all any AGN-$\nu$ expected connections. There are several persistent AGN flaring  TeV sources with no correlation with these 54 IceCube events.\\
4) There is a puzzling disagreement in power spectra among HESE events ($\sim E^{-2.6}$, similar to cosmic ray spectra) and the through-going ones ($\sim E^{-2.1}$), why?\\
5) There is a probable through-going event  leaving 4~PeV energy  released by a probable  8~PeV primary $\nu_\mu$ ; surprisingly, there is not yet any $\bar{\nu}_e+e\to W^-$ leading to Glashow resonant  event expected in HESE list at 6.3~PeV \cite{glashow}. This contradictions may become larger and larger because of the Glashow resonant absence event respect other additional PeVs signals.\\
6) No self-narrow clustering occurred within the UHE$\nu_\mu$ events. They are quite collimated signals, (contrary to cascade ones). There are indeed some weak correlation \cite{2015EPJWC..9908002F}, that might point to a few galactic sources, but at least a  more narrow clustering would favor a real true Astronomy.\\
7) The 3 years of IceCube record (2013-2015) are going to be doubled  soon (ICRC 2017, end of 2017$?$); additional volumes and improved techniques may double or tripled the event candidate to a tau appearance. If the $\nu_\tau$ ($\bar{\nu}_\tau$) will still be absent there will be just a main escape door: HESE events are mostly atmospheric prompt (charmed) events that are mostly made by $\nu_e$, $\bar{\nu}_e$, $\nu_\mu$, $\bar{\nu}_\mu$ in equal rate, with a negligible tau component \cite{2008Reno_TAU_PROMPT}. Their spectra then must be by definition a photocopy of Cosmic Rays one ($\sim E^{-2.6}$), as indeed they partially they show. The prompt neutrino asymmetric flavor spectra ($\phi_{\nu_\mu}:\phi_{\nu_e}:\phi_{\nu_\tau}=1:1:\frac{1}{20}$), disfavoring the tau flavor, as it has been observed by ICECUBE, it may oppose to the expected astrophysical democracy,(because complete mixing), of $\phi_{\nu_\mu}:\phi_{\nu_e}:\phi_{\nu_\tau}=1:1:1$ .\\
In this very first summary as we  underlined in the title, paraphrasing a famous Italian Martini spot, we are forced to claim once again ``No Tau? No Astronomy!''.

\begin{figure}[!t]
\begin{center}
\includegraphics[width=0.7\textwidth]{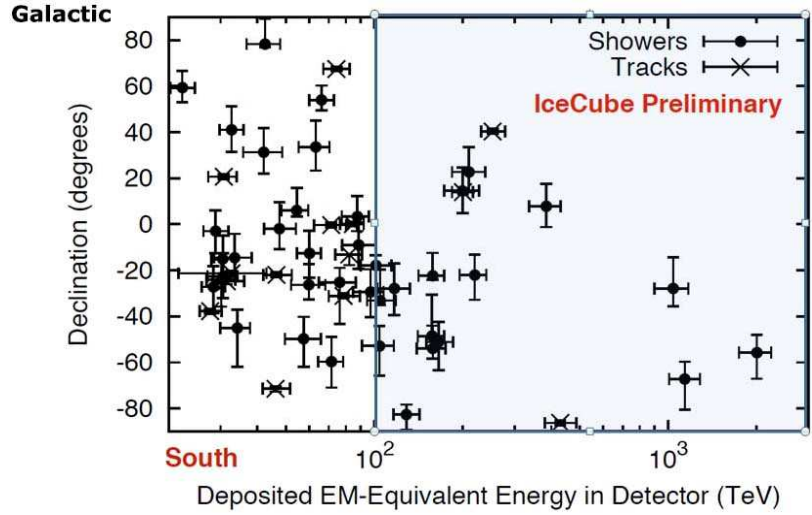}
\caption{The case of a future ability to verify the tau twin bangs also at 100 TeVs energy would be essential. For three years record there are nearly 18 events. In case of twice a record (at 2017-2018) the total number may be doubled up to 35 events.} \label{Fig3}
\end{center}
\end{figure}
\begin{figure}[!ht]
\begin{center}
\includegraphics[width=0.7\textwidth]{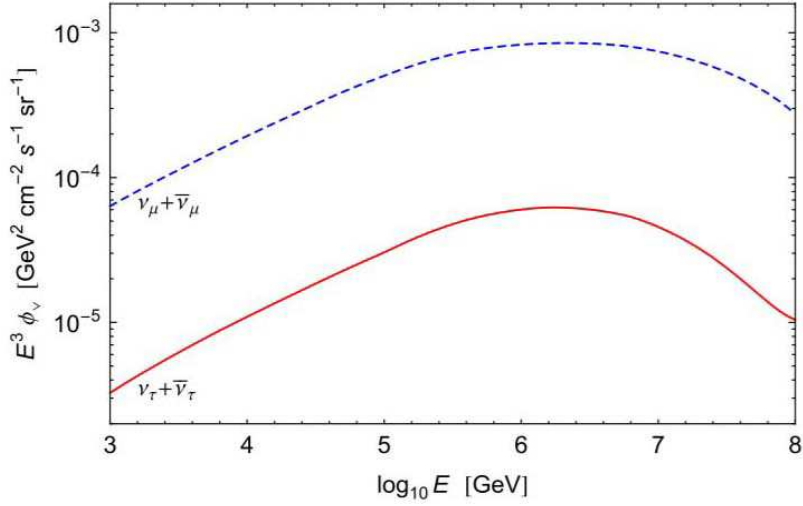}
\caption{In the simplest prompt neutrino spectra we must foresee the same rate for $\mu$ and $\nu_{\mu}$ as well as $\nu_{e}$ and their antiparticles, see \cite{2008Reno_TAU_PROMPT}. Anyway at lower rate nevertheless there must be also a tenth of tau signals.} \label{Fig4}
\end{center}
\end{figure}

\section{Near future tau detections}

The present and the future $\tau$ search may be soon improved by three factors:
The time larger sample; the enlarged IceCube area and volume; the lowering of the energy threshold from 200 TeV to 100 TeV.
Indeed a new method for astrophysical tau neutrino search in the
waveforms is being implemented by IceCube leading to a lower detection energy threshold to about (100 TeV).
The first time increase data imply that if for 914 days (three years) there have been recorded nearly ten events above 200 TeV, (see Fig. \ref{Fig1}), in 6 years data
that are being going to be possibly doubled.

\subsection{The simplest, the best: how many  tau double bang?}

Let us try to imagine the fate of an hadronic astrophysical neutrino:
If it has been originated by any common pion $\pi^{\pm}$ or $K^{\pm}$ decay,  first  into $\mu^{\pm}$ and later on by muon decay, we would expect  as a first approximation an astrophysical neutrino with no neutrino tau flavor $\ \Phi_{\nu_{e}}:\Phi_{\nu_{\mu}}:\Phi_{\nu_{\tau}}= 1:2:0$. Or for a normalized to unity flux: $\Phi_{\nu_{e}}:\Phi_{\nu_{\mu}}:\Phi_{\nu_{\tau}}= \frac{1}{3}:\frac{2}{3}:0 $.

 However the well known neutrino oscillation and mixing and final de-coherence will lead to an almost  complete democratic rate $\Phi_{\nu_{e}}:\Phi_{\nu_{\mu}}:\Phi_{\nu_{\tau}}=1:1:1$ or in normalized rate:
  $$\Phi_{\nu_{e}}:\Phi_{\nu_{\mu}}:\Phi_{\nu_{\tau}}= \frac{1}{3}:\frac{1}{3}:\frac{1}{3}$$
 The result is nearly the same for prompt (astrophysical) neutrino sources whose later on flight oscillations and mixing  will normalize in a democratic final rate. The consequence is the follow: if IceCube neutrinos are all astrophysical (as the highest ones at 200 TeV and above should be), than the expectance of their presence or absence is given by simple binomial distribution: in particular assuming an astrophysical neutrino where there is an equal probability to observe an electron, a tau or a muon or their  Neutral Current at equal $\frac{1}{4}$ rate,
   $$\Phi_{\nu_{e}}:\Phi_{\nu_{\mu}}:\Phi_{\nu_{\tau}}:\Phi_{\nu_{NC}}= \frac{1}{4}:\frac{1}{4}:\frac{1}{4}: \frac{1}{4}$$
  therefore the probability $P_{\mathrm{No}-\tau}$ not to observe any tau neutrino within N trials  is simply
\begin{equation}
P_{\mathrm{No}-\tau}=\left(\frac{3}{4}\right)^{N}.
\end{equation}
The consequent absence in the 914 days IceCube  old data (see Fig. \ref{Fig1}) of any tau signal among 9 or 10 events is respectively only:
\begin{equation}
 P_{\mathrm{No}-\tau}(9)=\left(\frac{3}{4}\right)^{9} = 7.5\%,\;\;\;P_{\mathrm{No}-\tau}(10)=\left(\frac{3}{4}\right)^{10} = 5.6\%.
\end{equation}
 These probabilities are already telling us that the tau absence is in tension with the astrophysical interpretation.
 If the same estimate will be doublet by twice the event number (see Fig. \ref{Fig3}) because a lower energy threshold at 100 TeV, or by a ten times larger IceCube 2 volume, we should wait for:
 \begin{equation}
 P_{\mathrm{No}-\tau}=\left(\frac{3}{4}\right)^{18} = 0.56\%.
\end{equation}
 This (a very near future hypothetical) case will exclude an astrophysical nature of IceCube events.
 On the same statistical view the larger acceptance for tau neutrino  in IceCube respect muon neutrino (see Fig. \ref{Fig2}) imply a contradictory statistical situation where among 10 events three are by muons (at least a factor $3.5$ suppressed by their cross section respect the tau one), none are tau.
 Indeed in this frame, assuming as obvious the equal NC (neutral current) contribute by all the three astrophysical neutrinos but keeping the muon suppression by a factor at least 3, we expect the following flux ratio of events:$\Phi_{\nu_{e}}:\Phi_{\nu_{\mu}}:\Phi_{\nu_{\tau}}:\Phi_{\nu_{NC}}=1:\frac{1}{3}:1:1$ that will reflect in a normalized

 $$\Phi_{\nu_{e}}:\Phi_{\nu_{\mu}}:\Phi_{\nu_{\tau}}:\Phi_{\nu_{NC}}= \frac{3}{10}:\frac{1}{10}:\frac{3}{10}: \frac{3}{10}$$.
 $$P_{\mathrm{No}-\tau}=\left(\frac{7}{10}\right)^{9} = 4\%.$$
 \begin{equation}
  P_{\mathrm{No}-\tau}=\left(\frac{7}{10}\right)^{10} = 2.82\%.
 \end{equation}
 In this prospective the ten times tau absent event imply already a significant tension with  present glorified astrophysical interpretation.
 These results will be greatly amplified by a doubled candidate number as twice the observed ones:
 \begin{equation}
  P_{\mathrm{No}-\tau}=\left(\frac{7}{10}\right)^{18} = 0.16\%.
 \end{equation}
 No tau appearance within 18 near future records may imply at $99.84\%$ level the inconsistence of the ICECUBE astrophysical neutrino origin.

 \subsection{The charmed neutrino flux and their tau suppression}
 Finally the alternative interpretation of charmed neutrino signals made (see Fig. \ref{Fig4}) by $\nu_{e}$, $\nu_{\mu}$ and by their antiparticles and by their common neutral current will imply a final  observed flux:
  \begin{equation}
\Phi_{\nu_{e}}:\Phi_{\nu_{\mu}}:\Phi_{\nu_{\tau}}:\Phi_{\nu_{NC}}=1:1:\frac{1}{10}:\frac{21}{30}
\end{equation}
 that last tiny term $\frac{1}{10}$ it is due to the additional $\nu_{\tau}$ rate, while the $\frac{21}{30}$  it is due to the electron, muon and small tau  neutral current events. In that case, as one may easily see, the absence of tau event it is a much more probable outcome; the normalized fluxes will be:
   \begin{equation}
\Phi_{\nu_{e}}:\Phi_{\nu_{\mu}}:\Phi_{\nu_{\tau}}:\Phi_{\nu_{NC}}=\frac{30}{84}:\frac{30}{84}:\frac{3}{84}:\frac{21}{84}
\end{equation}
The consequent probability to not observe a tau in charmed frame it is
\begin{equation}
  P_{\mathrm{No}-\tau}=\left(\frac{81}{84}\right)^{10} = 69.5\%
 \end{equation}
  that it is well compatible with the present IceCube tau absence.  In a near future, doubling the data,  this probability may becomes
  \begin{equation}
 P_{\mathrm{No}-\tau}=\left(\frac{81}{84}\right)^{18} = 51.96\%,\;  P_{\mathrm{No}-\tau}=\left(\frac{81}{84}\right)^{20} =  48.3\%.
  \end{equation}
    Therefore the eventual not discover or just a single tau discover  within next future 20 candidate events (above 200 TeV) may be (in equal probability) well allowed in the prompt atmospheric interpretation, while it will be still  a rare case, quite  well in tension with  any astrophysical interpretation:
  \begin{equation}
P_{\mathrm{No}-\tau}=\left(\frac{3}{4}\right)^{20} = 0.317\%,\;\;\;\; P_{\tau=1}=\left(\frac{3}{4}\right)^{19}\cdot \frac{1}{4}\cdot 20 = 2.11\%
 \end{equation}
The ideal case where the energy threshold is reduced to 100 TeV and the data record is doubled is leading to nearly 35 candidates.
 Therefore in this future optimal rate the probability to not observe a tau will be extremely severe
  \begin{equation}
 P_{\mathrm{No}-\tau}=\left(\frac{3}{4}\right)^{35} = 0.0042\%.
 \end{equation}

\section{Conclusion}
The main message of present article it is the following: there are already many hints and soon strong statistical signals
that the ICECUBE events might be mainly polluted by atmospheric neutrinos and very marginally (if any) by astrophysical ones.
The polluted atmospheric events are charmed ones at a rate just a little above the foreseen ones \cite{2008Reno_TAU_PROMPT}.

 Indeed the absence of tau signal in old three years record of IceCube data is allowed within a $P_{\mathrm{No}-\tau}= 5.6-7.5\%$ or better to say $P_{\mathrm{No}-\tau}= 4-2.8\%$  narrow windows of possibility assuming an astrophysical interpretation, while it is a quite  probable outcome in a main prompt charmed neutrino scenario $P_{\mathrm{No}-\tau}= 69.5\%$. The very near double future $18-20$ candidate events above 200 TeV in IceCube for the last 6-7 years record, the eventual persistent absence of the tau it will lead to
     $P_{\mathrm{No}-\tau}= 99.7\%$ or at $P_{\mathrm{No}-\tau}= 98\%$  probability, assuming that IceCube events are  of astrophysical nature. The probability to not observe them in atmospheric charmed scenario it is still very large.
      To accept the astrophysical neutrino nature we should  expect two or even more tau events within the next future 20 candidates. For instance the probability to not observe at least two tau in astrophysical vision within 20 events is:
 \begin{equation}
  P_{N_{\tau}=2}=\left(\frac{3}{4}\right)^{18}\cdot\left(\frac{1}{4}\right)^{2}\cdot 10\cdot19= 6.7\%.
 \end{equation}
 Thus, in a near future either we should observe several tau double bang, otherwise the neutrino astronomy is at stake, more or less severe. The consequence it is that any Neutrino astronomy it is below the present charmed atmospheric noise signal. Only and mainly the well filtered and astrophysical tau air-shower road-map \cite{2002-2004_Fargion} it might open in a guaranteed way our eyes to a neutrino sky, as in  present and future experiments \cite{PAO(2008)}, \cite{ASHRA(2013)}, \cite{GIANT(2017)}.


\begin{thebibliography}{99}
\bibitem{Fargion_2014}  Fargion, D.,Paggi,P. NIMA Volume 753, 21 July 2014, Pages 9--13 (2014)
\bibitem{Fargion_WHY} Fargion, D., Oliva, P. , Graziano U., PoS FRAPWS2014,028,(2016)
\bibitem{2017arXiv170200021F} Fargion, D., \& Oliva, P.\, arXiv:1702.00021, (2017)
\bibitem{2017arXiv170500383K} Kevin J.~Meagher on behalf of the IceCube Collaboration, arXiv:1705.00383,(2017)
\bibitem{2015EPJWC..9908002F} Fargion, D., Ucci, G., Oliva, P., \& De Sanctis Lucentini, P.~G.\ European Physical Journal Web of Conferences, 99, 08002 (2015)
\bibitem{2014NuPhS.256..213F} Fargion, D., Oliva, P., \& De Sanctis Lucentini, P.~G.\ 2014, Nuclear Physics B Proceedings Supplements, 256, 213 (2014)
\bibitem{2013arXiv1309.7003I} IceCube Collaboration, Aartsen, M.~G., Abbasi, R., et al., arXiv:1309.7003 (2013)
\bibitem{2016arXiv1605.00177} Fargion, D., Oliva, P., arXiv:1605.00177 (2017)
\bibitem{2017arXiv170502383I} IceCube Collaboration, Aartsen, M.~G., Ackermann, M., et al.\ 2017, arXiv:1705.02383, (2017)
\bibitem{1995APh.....3..267L} Learned, J.~G., \& Pakvasa, S.\ 1995, Astroparticle Physics, 3, 267, (1995)
\bibitem{2008Reno_TAU_PROMPT} Enberg,R.,  Reno,M.H., Sarcevic, I., Physical Rev. D 78, 043005 (2008)
\bibitem{2016PhRvD..93b2001A} Aartsen, M.~G., Abraham, K., Ackermann, M., et al.\ 2016, Phys. Rev. D 93, 022001,(2016)
\bibitem{2017arXiv170205238X} Xu, D., \& for the IceCube Collaboration 2017, PoS ICHEP2016, 452, arXiv:1702.05238 (2016)
\bibitem{2002-2004_Fargion} Fargion D., The Astrophysical Journal, 570:909-925 (2002); Fargion D. et al. The Astrophysical Journal, 613:12851301; (2004)
\bibitem{PAO(2008)}  Bluemer J., Pier Auger Collaboration, 10.1143/JPSJS.78SA.114 (2008)
\bibitem{ASHRA(2013)}Asaoka, Y. ,  Sasaki,M.,  ASHRA Collaboration, Astr Phys.  \textbf{41} 716, (2013)
\bibitem{glashow} Glashow, S. L. \ 1960, Phys. Rev. 118, 316 (1960)
\bibitem{GIANT(2017)} Martineau-Huynh O. et al. EPJ Web of Conferences 135, 02001 (2017)









%
%
%
%
%
%
%
%
%
%
%
%
%
%
%


%
%
%
%
%
%
%
\end{thebibliography}
\end{document}